\documentclass[amsmath,amssymb, showpacs,aps,notitlepage]{revtex4-1}
\usepackage{caption}
\usepackage{graphicx}% Include figure files
\usepackage{dcolumn}% Align table columns on decimal point
\usepackage{bm}% bold math
\begin{document}

\title{Petrov type of linearly perturbed type D spacetimes}

\author{Bernardo Araneda}
\author{Gustavo Dotti}
 \email{Electronic address: \texttt{gdotti@famaf.unc.edu.ar}}

\affiliation{ Facultad de Matem\'atica, Astronom\'{\i}a y F\'{\i}sica\\
 Universidad Nacional de C\'ordoba;\\
Instituto de F\'{\i}sica Enrique Gaviola, CONICET\\
Ciudad Universitaria, (5000) C\'ordoba, Argentina}

\date{\today}
\begin{abstract}
We show that a spacetime satisfying the linearized vacuum Einstein equations around a   type D 
background is generically  of type I, and that the splittings of the Principal Null Directions (PNDs) 
and of the degenerate eigenvalue of the Weyl tensor are non analytic functions of the perturbation 
parameter of the metric. This provides a gauge invariant characterization of the effect of the 
perturbation on the underlying geometry, without appealing to differential curvature invariants. This is 
 of particular interest for  the Schwarzschild solution, for which there are no
signatures of the even perturbations on the algebraic curvature invariants. We also show that, unlike the general 
case, the unstable even modes of the Schwarzschild naked singularity deforms the Weyl tensor 
into a type II one.
\end{abstract}

\maketitle

%\tableofcontents

\section{Introduction}

Understanding how the background geometry is affected by gravitational waves, here meaning  solutions 
 $h_{ab}$ to the linearized vacuum Einstein's equation 
(LEE) around a vacuum solution  $g_{ab}$: 
$$\lim_{\epsilon \to 0} \frac{R_{ab}[g_{cd}+ \epsilon h_{cd}]}{\epsilon}=0, $$
or, more explicitly,
\begin{equation} \label{lee}
 \nabla^{c} \nabla_{c} h_{a b} + \nabla_{a} \nabla_{b} (g^{c d} h_{c d}) -
 2 \nabla^{c} \nabla_{(a} h_{b) c} =0,
\end{equation}
 is a rather intricate problem, partly due to the gauge issue of linearized gravity, that is,  the fact that given a solution  $h_{a b}$ of 
(\ref{lee}) and an arbitrary vector field $V^c$, 
 $h'_{a b}$ defined by 
 \begin{equation} \label{gt}
h'_{a b} = h_{a b} +  \pounds_V   g_{a b},
\end{equation}
is  also solution, although physically equivalent to $h_{ab}$. We are only interested in the 
equivalence classes of solutions of (\ref{lee}) under 
the equivalence relation 
$h'_{a b} \sim h_{a b} +  \pounds_V   g_{a b}$,  and on functionals of $h_{ab}$ that are gauge invariant, i.e., depend only on the
 equivalence class of 
 $h_{ab}$. 
The linear perturbation $\delta T$ of a tensor field $T$  that is a functional of the metric, transforms as $\delta T \to \delta T +  
\pounds_V T$ under the 
gauge transformation (\ref{gt}), therefore only constant scalar fields and tensor products of the identity map $\delta^a_b$ are gauge 
invariants. \\
A possibility explored in  \cite{Dotti:2013uxa}  is to parametrize the equivalent classes of metric perturbations in terms of the perturbations  
of a 
 set of curvature  scalar fields that vanish in the background. In vacuum, there are only four functionally 
independent {\em algebraic} invariants of the Weyl tensor $C_{abcd}$ and its dual $C_{abcd} ^*$ (recall that left and right Hodge duals of 
$C_{abcd}$ agree),
these are
\begin{align} \label{einvs}
Q_+ &:=  \tfrac{1}{48}  C_{abcd} C^{abcd} \\ \nonumber
C_+ & := \tfrac{1}{96}  C_{ab}{}^{cd} C_{cd}{}^{ef} C_{ef}{}^{ab}, \\ \nonumber
Q_- &:=  \tfrac{1}{48}  C_{abcd} ^*  C^{abcd},\\ \nonumber
C_- &:= \tfrac{1}{96}  C_{ab}^* {}^{cd}  C_{cd}{}^{ef} C_{ef}{}^{ab}.
 \end{align}
For the Schwarzschild black hole only  $Q_-$ and $C_-$ vanish in the background, 
and its is shown in \cite{Dotti:2013uxa} that $\delta Q_-$ parametrizes the space 
of odd (also called {\em vector}, see, e.g., \cite{Sarbach:2001qq} \cite{Ishibashi:2011ws}) metric perturbations, and that 
 $\delta C_- \propto
\delta Q_-$ and therefore 
adds no information. On the other hand, under even (or {\em scalar},  \cite{Sarbach:2001qq} \cite{Ishibashi:2011ws})
 perturbations, every gauge invariant combination of the perturbed  scalars in (\ref{einvs})
vanishes identically.
 For this reason reason, the gauge invariant combination $(9M-4r) \delta Q_+ + 3r^3 \delta X$, which involves the 
{\em differential invariant}  $X= (1/720) (\nabla_a C_{bcde}) (\nabla^a C_{bcde})$  was added to $\delta Q_-$ 
in \cite{Dotti:2013uxa}  to parametrize the entire set of perturbations using geometrically meaningful quantities.
A natural question to ask is what -if any- are the effects of the even perturbations on the curvature itself; in other words, do we
really need to look at differential invariants to find a geometric signature of the perturbation? That contractions 
of products of the curvature tensor 
hide vital information in Lorentzian geometry is not a surprise: pp-waves are an extreme example of  non flat vacuum metrics
for which every scalar made out of $C_{abcd}$ and arbitrary covariant derivatives of it vanish \cite{Pravda:2002us}.\\

The Weyl tensor of a generic metric is of type-I in the Petrov classification. This means that the eigenvalue problem 
\begin{equation} \label{ev-i}
\tfrac{1}{2} C^{ab}{}_{cd} X^{cd} = \lambda X^{ab}, \;\; X^{ab}=X^{[ab]}
\end{equation}
admits three {\em different} solutions, with $\lambda_1+\lambda_2+\lambda_3=0$ or, equivalently, that the equation 
\begin{equation}\label{pnd-I}
 k_{[e} C_{a]bc[d}k_{f]}k^{b}k^{c}=0, \;\;\; k^a k_a=0,
\end{equation}
admits four solutions spanning four different null {\em lines} (called principal null {\em directions} or PNDs).
Type D spacetimes, instead, are characterized by the fact that the eigenvalue equation (\ref{ev-i}) admits three linearly independent
 solutions with $\lambda_1=\lambda_2=-\tfrac{1}{2} \lambda_3$, 
a condition that turns out to be equivalent to the existence of two  so called double PNDs, that is, two non proportional null vectors 
satisfying 
an equation stronger than (\ref{pnd-I}):
\begin{equation}\label{pnd-d}
 C_{abc[d}k_{e]}k^{b}k^{c}=0, \;\;\; k^a k_a=0.
\end{equation}
It should be stressed that  equations (\ref{pnd-I}) and (\ref{pnd-d}), being homogeneous, do not define (null) 
tangent vectors at a point $p$ of the spacetime (i.e., elements of $T_pM$), but instead one dimensional subspaces  through the origin of 
$T_pM$ (this is why se talk of null {\em directions}), 
that is, points in the projective space $P(T_pM)$. We may think that type I (D) spacetimes have four  (two) smooth valued functions that assign 
to every $p \in M$ a point in $P(T_pM)$.
Now suppose that $g_{ab}(\epsilon)$ ($\epsilon$ in an open interval around zero) 
is a mono-parametric family  of vacuum solutions  of type I for $\epsilon \neq 0$ 
and type D for $\epsilon =0$. The tensor 
\begin{equation} \label{h}
h_{ab} = \lim_{\epsilon\to 0} \frac{g_{ab}(\epsilon)-g_{ab}(0)}{\epsilon},
\end{equation}
satisfies (\ref{lee}) and 
the four PNDs of $C_{abcd}(\epsilon)$ coalesce pairwise into two PNDs as $\epsilon \to  0$.
We can choose solutions $k_1^a(\epsilon), k_2^a(\epsilon), k_3^a(\epsilon)$ and $k_4^a(\epsilon)$ of 
$k_{[e} C_{a]bc[d}(\epsilon)k_{f]}k^{b}k^{c}=0$ such that $k_1(0)=k_2(0)$ and $k_3(0)=k_4(0)$ 
are the two non-proportional solutions of the type-D equation $ C_{abc[d}(0)k_{e]}k^{b}k^{c}=0$. 
Similarly 
we can label the eigenvalues  $\lambda_1(\epsilon), \lambda_2(\epsilon)$ and $\lambda_3(\epsilon)$ of (\ref{pnd-I}) such that 
$\lambda_1(0)=\lambda_2(0)=-\tfrac{1}{2} \lambda_3(0)$. This suggests using either the splitting of  type D PNDs in $P(T_pM)$
 into a pair of PNDs, or 
the eigenvalue splitting
\begin{equation}\label{sign}
\pm (\lambda_2(\epsilon)-\lambda_1(\epsilon))
\end{equation}
to measure  the distortion of the unperturbed $\epsilon=0$ metric. Note that the sign ambiguity in (\ref{sign}) 
comes from the $1 \leftrightarrow 2$ freedom in labeling the two eigen-bivectors (\ref{ev-i}) that degenerate in the $\epsilon=0$ limit.
 We have found that for  $\epsilon \simeq 0$, these are  
 appropriate, gauge invariant estimates of the distortion of the geometry by a gravitational wave involving only the Weyl tensor.\\

The reason why we expect these quantities to be non trivial is that the bulk of vacuum solutions of the Einstein's equation are type-I,
 while algebraic
special solutions comprise a zero measure subset ${\cal A}$. The curve $g_{ab}(\epsilon)$ will generically be transverse to ${\cal A}$ at
 $g_{ab}(0)$ (as assumed above), 
its ``tangent vector'' $h_{ab}$ will stick out of ${\cal A}$ (more on this in Section \ref{asp}). \\

An {\em speciality index} $\mathcal{S}$ was introduced in \cite{camp1} based on the fact that a vacuum spacetime is algebrically special 
if and only if $C_+^2=Q_+^3$; it was defined as $\mathcal{S}:=C_+^2/Q_+^3$ and the departure of this index from unity  is  regarded  as a measure 
of the ``degree of non-speciality''  of a metric. Although useful in numerical evolutions of the full Einstein equation, $\mathcal{S}-1$ 
vanishes identically in linear theory, suggesting the -wrong- conclusion that linearly perturbed algebraically special spacetimes remain special. 
This error was nicely clarified in the paper \cite{cherubini}, taking advantage of the fact that a mono-parametric set of   vacuum solutions 
within the Kasner family, with $g_{ab}(0)$ being type D, can be explicitly constructed. From the results in \cite{cherubini} is to be expected 
that the PND splitting   will be a power series of $\epsilon^{1/2}$. This non analyticity on $\epsilon$ comes from the fact that the associated algebraic problem
involve radicals that vanish for $\epsilon = 0$ \cite{cherubini}.\\

Stationary electro-vacuum black holes  have the algebraic symmetries of  type D. Deviations from these metrics
 represent a number of very
 important astrophysical processes and, as argued above,  will not to be algebraically special. 
 Thus, in the context of black hole perturbation theory it is important to study whether or not the
 perturbative techniques capture these algebraic
 aspects of the geometry. In numerical application this question has   been analyzed in \cite{camp1},\cite{camp2}.
 On the analytical side  in \cite{cherubini} for the Kasner type D solutions. 
In the present work we focus on the problem of finding expressions for the eigenvalue and PND splittings 
of the type D background under 
  \textit{arbitrary} linear perturbations. 

As an application, we  consider the effects of  linear perturbations of the Schwarzschild black hole solution and show that the  even or scalar 
gravitational waves (those that do not leave a visible trace in the perturbed algebraic curvature invariants) do 
affect the Weyl tensor by turning it into  type I, and that this effect is accounted for by the proposed gauge invariants above and 
we also show that these invariants mix  harmonic modes. 
As far as we are aware, these 
signatures of the black hole linear perturbation are analyzed in this form for the first time here. \\
Kerr black holes and Chandrasekhar algebraic special modes in Schwarzschild naked singularities are also briefly considered.

\section{Petrov types}

Although  the simplest approach to the Petrov classification is accomplished  using spinor methods, perturbation theory is much more 
tractable  in tensorial language, which is the one we adopt in this paper. Newman-Penrose equations will also be avoided, although 
complex methods and, in particular, a complex null tetrad will be used. For the sake of completeness, and to clarify some aspects 
of the linearized theory calculations, we briefly review the eigenvalue and PNDs approaches to the Petrov classification 
in this section.

\subsection{Eigenvalue approach}

The eigenvalue  approach to the Petrov classification regards the Weyl tensor as a linear map $X^{cd} \to C^{ab}{}_{cd} X^{cd}$ in the 
space of rank two 
 antisymmetric tensors (also called ``bivectors''), and analyzes the eigenvalue equation (\ref{ev-i}). 
The six dimensional bivector space is real isomorphic to the complex three dimensional  space of self-dual bivectors (SDB), 
which are those satisfying ${}^*S_{ab}:=\tfrac{1}{2} \epsilon^{ab}{}_{cd} S^{cd} = -i S^{ab}$. The isomorphism is given by
\begin{equation} \label{x2sd}
 X^{ab} \to X^{ab} + \tfrac{i}{2}  \epsilon^{ab}{}_{cd} X^{cd}  =: \tilde X ^{ab},
\end{equation}
its inverse being 
\begin{equation} 
X^{ab} =  \tfrac{1}{2} \left(  \tilde X ^{ab} + c.c. \right)
\end{equation}
The space of SDBs, in turn, is isomorphic to $u_{\perp}^{\mathbb C}$,  the complexification of the space of vectors orthogonal to a
 given   unit time-like vector $u^a$. The isomorphism is 
\begin{equation} \label{xvec}
\tilde X^{ab} \to \tilde X^{ab} u_b =: X^a
\end{equation}
and its inverse is
\begin{equation} \label{inverse}
\tilde X^{ab} = 2 u^{[a} X^{b]} + i \epsilon^{ab}{}_{cd} u^c X^d.
\end{equation}
If $\Lambda^a{}_b$ is a Lorentz transformation of unit determinant
\begin{equation} \label{lt}
\Lambda^a{}_c \Lambda^b{}_d g^{bd} = g^{ab}, \;\; 
\Lambda^a{}_p \Lambda^b{}_q\Lambda^c{}_r \Lambda^d{}_s \epsilon^{pqrs} = \epsilon_{abcd},
\end{equation}
 one obtains from  (\ref{lt}) that $(\Lambda^{-1})^a{}_b = \Lambda_b{}^a$ and 
\begin{equation} \label{c}
\Lambda^a{}_p \Lambda^b{}_q \epsilon^{pq}{}_{cd} = \epsilon^{ab}{}_{rs} \Lambda^r{}_c \Lambda^s{}_d.
\end{equation}
Equation (\ref{c})  implies that the unit determinant Lorentz transformations commute with  the map (\ref{x2sd}), as expected  from 
\begin{equation}
\tfrac{1}{2} X^{ab} X_{ab} = \tilde{X}^{ab} \tilde X_{ab} = - 4X^a X_a.
\end{equation}
The second equality above, obtained from (\ref{inverse}) 
 together with $u_a X^a =0$,  indicates that $\Lambda$ acts on $u_{\perp}^{\mathbb C}$ as an $SO(3,\mathbb{C})$ transformation 
$X^a \to A^a{}_b X^b$. With the help of (\ref{inverse}) and (\ref{xvec}) we find that this gives an
 isomorphism between 
$SO(3,1)^{\uparrow}$ (the group of unit determinant Lorentz transformation preserving time orientation, i.e., 
$\Lambda^a{}_b u_a u^b < 0$ 
if $u^c$ is time-like) and $SO(3,{\mathbb C})$:\\
\begin{equation} \label{so3}
A^a{}_k =  \Lambda^0{}_0 \; \Lambda^a{}_k -\Lambda^a{}_0 \; \Lambda^0{}_k + i \epsilon^{0 a}{}_{ql} \: \Lambda^q{}_0 \; \Lambda^l{}_k,
\end{equation}
where a zero down (up)  index means contraction with $u^b$ ($-u_b$). 
To gain some intuition on the isomorphisms (\ref{x2sd}) and (\ref{xvec}),
 note that if $X_{ab}$ is the electromagnetic tensor, then $X^a = E^a + i B^a$, the electric and magnetic 
fields measured by an observer moving with  velocity $u^a$. If $\Lambda^a{}_b u^b = u^a$, then $\Lambda$ is a rotation, and (\ref{so3}) reduces to 
\begin{equation} \label{so3r}
A^a{}_k = u^a{} \; u_k +  \Lambda^a{}_k 
\end{equation}
which belongs to $SO(3,{\mathbb R})$ and simply rotates $\vec{E}$ and $\vec{B}$ independently. Otherwise, $\Lambda$ is a boost and 
(\ref{so3})  a complex rotation 
mixing $\vec{E}$ and $\vec{B}$, as expected. \\
We can replicate the above constructions for the self-dual piece of the Weyl tensor $C_{abcd}$ by regarding it as 
an element of the symmetric tensor product of bivector space,  and using the fact that that left self-duality implies right self duality. Define 
\begin{equation} \label{sdw}
\tilde C_{abcd} := \tfrac{1}{2} \left( C_{abcd} + \tfrac{i}{2} \epsilon_{ab}{}^{kl} C_{klcd} \right), 
\end{equation}
and introduce the  map $Q: u_{\perp}^{\mathbb C} \to u_{\perp}^{\mathbb C}$
\begin{equation}
Q^a{}_c := -\tilde C^a{}_{bcd} \; u^b \;  u^d.
\end{equation}
The above equation can again be inverted  \cite{stephani},
\begin{equation}
-\tfrac{1}{2} \tilde{C}_{abcd} = 4 u_{[a} Q_{b] [d}u_{c]} + g_{a [c} Q_{d] b} -  g_{b [c} Q_{d] a}
+ i \epsilon_{abef} u^e u_{[c}Q_{d]}{}^f + i \epsilon_{cdef} u^e u_{[a}Q_{b]}{}^f, 
\end{equation}
and the eigenvalue problem (\ref{ev-i}) is easily seen to be equivalent to 
$\tfrac{1}{4} \tilde C^{ab}{}_{cd} \tilde X^{cd} = \lambda \tilde X^{ab}$, and also to 
\begin{equation} \label{ev-ii}
Q^a{}_b X^b = \lambda X^a, \;\; X^a = \tilde X^{ab} u_b.
\end{equation}
We should stress here that, although $Q^a{}_b$ and $ X^b$ were defined using a particular unit timelike vector $u^c$, 
the eigenvalue equation (\ref{ev-ii}), being equivalent to (\ref{ev-i}), gives  covariant information and therefore is fully meaningful. 
Let $\{ e_o^a=u^a, e_1^a,e_2^a,e_3^a \}$ be an orthonormal tetrad  ($g_{ab} e^a_{\alpha} e^b_{\beta}= \text{diag} (-1,1,1,1)$)
 adapted to $u^a$, 
\begin{equation} \label{nt}
k^a = \frac{e_0^a+ e_3^a}{\sqrt{2}}, \;\;\; l^a= \frac{e_0^a- e_3^a}{\sqrt{2}}, \;\;\; m^a = \frac{e_1^a -i  e_2^a}{\sqrt{2}},
 \;\;\; \bar m ^a = \frac{e_1^a + i e_2^a}{\sqrt{2}}, 
\end{equation}
the related complex null tetrad. A basis of self-dual two forms is \cite{stephani}
\begin{equation} \label{sd}
U_{ab} = 2 \bar m_{[a} l_{b]}, \;\; V_{ab} = 2 k_{[a} m_{b]}, \;\; W_{ab} = 2 ( m_{[a} \bar m _{b]} + l_{[a} k_{b]} )
\end{equation}
 (note that the only non-zero contractions 
are $U_{ab} V^{ ab} =2$ and $W_{ab} W^{ ab} =-4$.) 
These can be used to expand 
\begin{equation} \label{w1}
\tfrac{1}{2} \tilde C = \Psi_0 UU + \Psi_1 (UW+WU) + \Psi_2 (VU+UV+WW ) + \Psi_3 (VW+WV) +\Psi_4 VV
\end{equation}
where $UV$ stands for $U_{ab} V_{cd}$, etc., and 
\begin{align}
\nonumber \Psi_0&:=C_{abcd}k^a m^b k^c m^d, \hspace{0.4cm} \Psi_1:=C_{abcd}k^a l^b k^c m^d \\
\nonumber \Psi_2&:=C_{abcd}k^a m^b \bar{m}^c l^d, \hspace{0.4cm} \Psi_3:=C_{abcd}k^a l^b \bar{m}^c l^d\\
\Psi_4&:=C_{abcd}\bar{m}^a l^b \bar{m}^c l^d. \label{w2}
\end{align}
 $ u_{\perp}^{\mathbb C}$ is the complex span of $\{e_1^a,e_2^a,e_3^a \}$ and, in this basis,  
\begin{equation}\label{Q}
Q^i{}_j=\left( \begin{matrix} -\frac{1}{2}\Psi_0+\Psi_2-\frac{1}{2}\Psi_4 & \frac{i}{2}(\Psi_4-\Psi_0) & \Psi_1-\Psi_3 \\
		\frac{i}{2}(\Psi_4-\Psi_0) & \frac{1}{2}\Psi_0+\Psi_2+\frac{1}{2}\Psi_4 & i(\Psi_1+\Psi_3) \\
		\Psi_1-\Psi_3 & i(\Psi_1+\Psi_3) & -2\Psi_2 \end{matrix} \right)
\end{equation}

If $Q$ has three distinct eigenvalues, then the algebraic type of the spacetime is  I. If instead
 two of them equal, say $\lambda_1=\lambda_2=:\lambda$, the space is of type II  
if $\text{dim(ker}(\textbf{Q}-\lambda\textbf{I}))=1$ or type D if $\text{dim(ker}(\textbf{Q}-\lambda\textbf{I}))=2$. 
Finally, in the
 case in which all of three eigenvalues are identical (then necessarily equal to zero, since $Q^i{}_i=0$),
 the Petrov type will be  be III, N or O, if
 $\text{dim(ker}(\textbf{Q}))=1$, $2$ or $3$, respectively. The matrix (\ref{Q}) representing $Q^i{}_j$ 
can be put in normal form (that is, diagonal or Jordan form) by acting on it 
with elements of  $SO(3,\mathbb{C})$. As $SO(3,\mathbb{C})$ is isomorphic to $SO(3,1)^{\uparrow}$ (c.f. equation (\ref{so3})),
 any such transformation is
 uniquely associated with a (proper,
 orthochronous) Lorentz
 transformation on the spacetime, and this transformation, acting on the null tetrad above, produces what 
is called a {\em principal tetrad}. The transformation 
leading to the normal form of $Q$, and therefore the principal tetrad,   is uniquely determined in the cases of Petrov type I, II and III 
(neither $k^a$ nor $l^a$ of
 the unique principal tetrad  give a PND in type I spacetimes \cite{hall}). For a type D space,
 however, there is a 2-dimensional residual $U(1) \times {\mathbb R}_{>0}{}^{\times}
= C^{\times}$ subgroup of $SO(3,1)^{\uparrow} \simeq SO(3,\mathbb{C})$ 
 of boost and rotations
preserving the normal form (and thus the PNDs):
\begin{equation} \label{sw}
k^a \to \alpha k^a, \;\;\; l^a \to \alpha^{-1} l^a, \;\;\; m^a \to e^{i \theta} m^a, \;\;\; \bar m^a \to e^{-i \theta} \bar m^a,
\end{equation}
In this case, 
and $k^a$ and $l^a$ are  aligned along the repetad PNDs, i.e., they satisfy (\ref{pnd-d}). Principal null tetrad components of tensors  
are said to carry spin-weight $s$ and boost-weight $q$  if under (\ref{sw}) they pick up a factor $e^{is \theta} \alpha^q$ (e.g., $\Psi_3$ 
has $s=q=-1$). Truly scalar fields,  such as $Q_+ \propto \Re( \Psi_2{}^2)$, of course, carry zero weights.

\subsection{Principal null directions}

An alternative approach to the Petrov classification consists of 
 studying the  PNDs of the Weyl tensor, i.e., solving equation (\ref{pnd-I}), which is equivalent to
\begin{equation}
k_{[e}\widetilde{C}_{a]bc[d}k_{f]}k^{b}k^{c}=0. \label{nulldir}
\end{equation}
Starting from  a generic null tetrad with associated Weyl scalars (\ref{w2}) we find that (see (\ref{w1}))
\begin{equation}
\tfrac{1}{2}k_{[e}\widetilde{C}_{a]bc[d}k_{f]}k^{b}k^{c}=\Psi_0 k_{[e}\bar{m}_{a]}\bar{m}_{[d}k_{f]},
\end{equation}
so the $k^a$ vector of the tetrad is a PND if an only the $\Psi_0$ component of the Weyl tensor in this tetrad vanishes
If we apply a null rotation (boost)  to the given null tetrad around  $l^a$, 
\begin{eqnarray} \nonumber
l^a &\rightarrow& l'{}^{a}=l^a,\\ \nonumber 
k^a &\rightarrow& k'^{a}=k^a+z\bar{z}l^a+\bar{z}m^a+z\bar{m}^a, \\
m^a &\rightarrow& m'^{a} = m^a+zl^a, \label{nr}
\end{eqnarray}
the resulting $k'^a$ will sweep the  $S^2$ set of null directions as $z$ moves in the complex plane, avoiding only the $l^a$ direction, 
wich corresponds to $z= \infty$ ($S^2=$ complex plane plus point at infinity). So we can calculate $\Psi_0'(z)$ in the primed tetrad 
and solve the fourth-order equation $\Psi_0'(z)=0$ to find the four PNDs. It can easily be checked using (\ref{w2}) that 
\begin{equation}
\Psi_0'(z) =\Psi_0+4z\Psi_1+6z^2\Psi_2+4z^3\Psi_3+z^4\Psi_4 \label{pol}
\end{equation}
 Generically (type I spaces), there will be four 
different solutions $z_j, j=1,2,3,4$ corresponding to four PNDs. The special cases are those for 
which the polynomial (\ref{pol}) has repeated roots, and can be classified according to the partitions of $4$ as 
\begin{center}
\begin{tabular}{c|c}
Petrov type & PNDs \\ \hline
I & $\{1111\}$ \\
II & $\{211\}$ \\
D & $\{22\}$ \\
III & $\{31\}$ \\
N & $\{4\}$ \\
O & $\{-\}$ 
\end{tabular}
\captionof{table}{Petrov type according to the multiplicity of the roots of the polynomial in (\ref{pol}).} 
\end{center}
In particular, for type D there are two double roots, and the corresponding $k^a$ will satisfy the stronger equation (\ref{pnd-d}). 
Type O corresponds to conformally flat spaces $C_{abcd}\equiv0$.

\section{Linear perturbations}

Let $g_{ab}(\epsilon)$ be a monoparametric family of vacuum solutions with $g_{ab}(0)=:g_{ab}$ of type D.
 Assume $e_{\alpha}^a(\epsilon)$ is an orthonormal tetrad of the metric $g_{ab}(\epsilon)$, smooth in $\epsilon$, 
and such that the associated null tetrad (\ref{nt}) has $k^a(0)$ and $l^a(0)$ aligned along the two repeated PNDs of the 
type D background $g_{ab}$, i.e., they satisfy (\ref{pnd-d}). If $\Lambda(\epsilon)$ is a curve in $SO(3,1)^{\uparrow}$ with $\Lambda(0)$
the identity, then 
the tetrad $\tilde e^a_{\beta}(\epsilon):=\Lambda^{\alpha}{}_{\beta}(\epsilon) e^a_{\alpha}(\epsilon)$ satisfy this same condition 
 (this is sometimes called the ``tetrad-gauge
ambiguity''). In any case $ \Psi_0(\epsilon) = (1/4) C(\epsilon) V(\epsilon) V(\epsilon)$ (using (\ref{w1}) and an obvious notation) and 
\begin{equation} \label{dp0}
\delta \Psi_0 := \frac{d}{d \epsilon} {\Big|}_{\epsilon=0} \Psi_0 = \tfrac{1}{4} (\delta C \; V V + C \delta V V + C V \delta V) = \tfrac{1}{4} \delta C \; V V, 
\end{equation}
since $C \delta V V = C V \delta V = \Psi_2 \delta V_{ab} V^{ab} =0$. Equation (\ref{dp0}) implies that 
$\delta \Psi_0$ is {\em tetrad-gauge invariant}. The reader can check that $\delta \Psi_2$ and $\delta \Psi_4$ are also  tetrad-gauge invariant, with 
\begin{equation} \label{dp4}
\delta \Psi_4 = \tfrac{1}{4} \delta C \; U U. 
\end{equation}
Note from (\ref{sd}), (\ref{sw}), (\ref{dp0}) and (\ref{dp4}) that $\delta \Psi_0$ ($\delta \Psi_4$) has spin weight two and 
boost weight two (minus two and minus two respectively).

\subsection{Perturbed eigenvalues}

To first order in $\epsilon$:
\begin{equation}\label{psi}
\Psi_{2}(\epsilon)=\Psi_2+\epsilon\delta\Psi_2, \hspace{0.5cm} \Psi_i(\epsilon)=\epsilon\delta\Psi_i, \:\: i=0,1,3,4 
\end{equation}
Inserting this in (\ref{Q}) we find that the perturbed eigenvalues to order $\epsilon$ are 
\begin{eqnarray}
\lambda_1(\epsilon)&=&\Psi_2+(\delta\Psi_2-\sqrt{\delta\Psi_0\delta\Psi_4})\epsilon,\\
\lambda_2(\epsilon)&=&\Psi_2+(\delta\Psi_2+\sqrt{\delta\Psi_0\delta\Psi_4})\epsilon,\\
\lambda_3(\epsilon)&=&-2\Psi_2-2\delta\Psi_2\epsilon,
\end{eqnarray}
and the eigenvalue splitting (\ref{sign}) is 
\begin{equation} \label{split}
2 \epsilon \; \sqrt{\delta\Psi_0\delta\Psi_4} , 
\end{equation}
the branch choice of the (complex) square root being responsible of the  sign ambiguity anticipated in (\ref{sign}). It is important to 
emphasize  
that $\delta\Psi_0$ and $\delta\Psi_4$  are booth free of the tetrad-gauge ambiguity and that they 
carry opposite spin and boost weights (see the discussion around
equation (\ref{sw})). Thus (\ref{split}) is a  well defined scalar field that carries information on the 
distortion of the Weyl tensor due to the perturbation, 
information that is missing, e.g.,  by the perturbed curvature scalars in the even sector of the Schwarzschild perturbations.\\
 If either $\delta\Psi_0=0$ or $\delta\Psi_4=0$, the space 
degenerates into a type D or II, depending on the dimension of the eigenspace ker$(\textbf{Q}-\lambda_2\textbf{I})$. 
We will comment on these {\em algebraically special perturbations} in section \ref{asp}.

\subsection{Splitting of the PNDs}

Replacing (\ref{psi}) in (\ref{pol}) gives
\begin{equation}
P(z):=\delta\Psi_0\epsilon+4\delta\Psi_1\epsilon z+6(\Psi_2+\epsilon\delta\Psi_2)z^2+4\delta\Psi_3\epsilon z^3+\delta\Psi_4\epsilon z^4.
\end{equation}
The equation to be solved is $P(z)=0$ up to order $\epsilon$.
 Note, however,  that the solutions $z_{\pm}$ of the simpler equation  $0=\delta\Psi_0\epsilon+6\Psi_2 z^2$,
\begin{equation}
z_{\pm}=\pm\sqrt{-\frac{\delta\Psi_0}{6\Psi_2}}\sqrt{\epsilon}, \label{roots}
\end{equation}
satisfy
\begin{equation}
P(z_{\pm})=0+\mathcal{O}(\epsilon^{3/2}),
\end{equation}
i.e., they are (to order $\epsilon$) two of the four solutions of $P(z)=0$.
Since $z_{\pm} \to 0$ with $\epsilon$, these two solutions  are easily guessed to be those related to 
the splitting of $k^a$ into two different PNDs.  Explicitly, in the dominant order we have
\begin{equation} \label{ks}
k^a_{\pm}(\epsilon):= k^a\pm
\epsilon^{1/2}\left[\sqrt{-\frac{\delta\Psi_0}{6\Psi_2}}\bar{m}^a+\overline{\left(\sqrt{-\frac{\delta\Psi_0}{6\Psi_2}}\right)}m^a \right]. 
\end{equation}
According to the discussion between equations (\ref{nt}) and (\ref{pol}) the other two solutions of $P(z)=0$
should be near the unperturbed repeated PND $l^a$,  which corresponds to $z=\infty$ 
 in  $S^2 = {\mathbb C} \cup \{ \infty \}$, thus we
 expect the other two 
solutions  to behave as an 
inverse power or $\epsilon$ (c.f. \cite{cherubini}). To obtain these, we
 can either switch to $x=1/z$, or work with null rotations around $l^a$ and
 solve the equation 
$\Psi_4=0$. In either case we arrive at 
\begin{equation} \label{ls}
l^a_{\pm}(\epsilon):=l^a\pm\epsilon^{1/2}\left[\overline{\left(\sqrt{-\frac{\delta\Psi_4}
{6\Psi_2}}\right)}\bar{m}^a+\sqrt{-\frac{\delta\Psi_4}{6\Psi_2}}m^a  \right] 
\end{equation}
It is important to note that (\ref{ks}) has zero spin weight and boost weight one, and thus define a  PND, for which 
 the overall scaling is 
 irrelevant. Similarly 
 (\ref{ls}) carries zero spin weight and boost weight minus one. The non-analytical character of the splitting, discussed in 
some detail in \cite{cherubini}, can be avoided by a re-parametrization of $g_{ab}(\epsilon)$.

\section{Applications}

\subsection{Gravitational perturbations  of the Schwarzschild black hole}

For the Schwarzschild solution 
\begin{equation} \label{schwa}
ds^2 = -f \; dt^2 + \frac{dr^2}{f}+ r^2 \left( d\theta^2 + \sin^2 \theta \; d\varphi^2 \right), \;\; f = 1 -\tfrac{2M}{r}, 
\end{equation}
we use the orthonormal tetrad
\begin{equation}
e_0^{a}=\frac{1}{\sqrt{f}}\left(\frac{\partial}{\partial t}\right)^a, \hspace{0.5cm} 
e_3^{a}=\sqrt{f}\left(\frac{\partial}{\partial r}\right)^a, 
\hspace{0.5cm} e_1^{a}=\frac{1}{r}\left(\frac{\partial}{\partial \theta}\right)^a, \hspace{0.5cm} 
e_2^{a}=-\frac{1}{r\sin\theta}\left(\frac{\partial}{\partial \phi}\right)^a,
\end{equation}
then $k^a$ and $l^a$ in (\ref{nt}) are along the repeated PNDs. 
The perturbed metric admits a series expansion  using a basis constructed from  harmonic tensors for
 $S^2$ \cite{Sarbach:2001qq,Ishibashi:2011ws},  labeled $(\ell,m,\pm)$, which  can be obtained by applying a differential 
operator to the standard 
spherical harmonic scalars $Y^{(\ell,m)}$ and are classified into even (+, or scalar) and odd (-, or vector) types 
 according to their behavior  under the discrete parity isometry $(\theta,\varphi)\rightarrow(\pi-\theta,\varphi+\pi)$
The linearized Einstein's equations reduce to  two-dimensional wave equations  for 
  functions $\phi_{(\ell,m)}^{\pm}(t,r)$ (see \cite{Sarbach:2001qq,Ishibashi:2011ws}), called the Regge-Wheeler and Zerilli equations.
In terms of these potentials 
we have found that 
\begin{align} \label{dp0s}
\delta\Psi_0 &=\sum_{(\ell,m,\pm)} A_{(\ell,m)}^{\pm}(t,r) \; Y_{2}^{(\ell,m)}(\theta,\phi) ,\\
\delta\Psi_4 &=\sum_{(\ell,m,\pm)} B_{(\ell,m)}^{\pm}(t,r) \; Y_{-2}^{(\ell,m)}(\theta,\phi)  \label{dp4s},
\end{align}
where, 
\begin{align} \nonumber
A_{(\ell,m)}^- &= -\frac{3iM}{8r^3} \sqrt{\frac{(\ell+2)!}{(\ell-2)!}} \left[(M-r)\frac{\partial\phi}{\partial r}^--r\left(\frac{r-3M}{r-2M}\right)\frac{\partial\phi}{\partial t}^- 
+r(2M-r)\frac{\partial^2\phi}{\partial r^2}^- 
-r^2\frac{\partial^2\phi}{\partial t\partial r}^-+\left(\frac{\ell(\ell+1)}{2}-\frac{3M}{r} \right)\phi^- \right]  \\
A_{(\ell,m)}^+ &=  - \sqrt{\frac{(\ell+2)!}{(\ell-2)!}} \left[ \frac{2M-r}{2r^2}\; \frac{\partial^2\phi}{\partial r^2}^+ -\frac{1}{2r} \frac{\partial^2\phi}{\partial r \partial t}^+
+\frac{K(r)}{2r^3 D(r)}  \frac{\partial\phi}{\partial r}^++ \frac{L(r)}{2 r^2 (r-2M) D(r)} \frac{\partial\phi}{\partial t}^++
\frac{N(r)}{4 r^4 D(r)} \phi^+\right],\\ \nonumber
B_{(\ell,m)}^- &=  -\frac{3iM}{8r^3} \sqrt{\frac{(\ell+2)!}{(\ell-2)!}} 
\left[(M-r)\frac{\partial\phi}{\partial r}^-+r\left(\frac{r-3M}{r-2M}\right)\frac{\partial\phi}{\partial t}^- 
+r(2M-r)\frac{\partial^2\phi}{\partial r^2}^-
+r^2\frac{\partial^2\phi}{\partial t\partial r}^-+\left(\frac{\ell(\ell+1)}{2}-\frac{3M}{r} \right)\phi^- \right] \\
B_{(\ell,m)}^+&=  - \sqrt{\frac{(\ell+2)!}{(\ell-2)!}} \left[\frac{2M-r}{2r^2}\; \frac{\partial^2\phi^+}{\partial r^2} +
\frac{1}{2r} \frac{\partial^2\phi^+}{\partial r \partial t}
+\frac{K(r)}{2r^3 D(r)}  \frac{\partial\phi^+}{\partial r}- \frac{L(r)}{2 r^2 (r-2M) D(r)} \frac{\partial\phi^+}{\partial t}+
\frac{N(r)}{4 r^4 D(r)} \phi^+\right],
\end{align}
$\phi^{\pm}$ stands for $\phi^{\pm}_{(\ell,m)}$, 
 $Y_s^{(\ell,m)}$ are the normalized spin weight $s$ spherical harmonics on $S^2$ \cite{Sarbach:2001qq}, and 
\begin{align}
D(r) &= (\ell+2)(\ell-1)r + 6M \\
K(r) &= (\ell+2)(\ell-1)(M-r)r-6M^2 ,\\
L(r) &= (\ell+2)(\ell-1)(3M-r)r+6M^2 , \\
N(r) &=  (\ell+2)^2(\ell+1) \ell (\ell-1)^2 r^3 + 6M (\ell+2)^2 (\ell-1)^2 r^2 + 36M^2 (\ell+2)(\ell-1) r +72M^3.
\end{align}

Note that the eigenvalue splitting (\ref{split}), as well as the repeated PNDs splittings (\ref{ks}) and (\ref{ls}), being 
proportional to $\sqrt{\Psi_0}$ and/or $\sqrt{\Psi_4}$ will contain multiple harmonics even  if the metric perturbation contains 
a single non-zero $\phi_{(\ell,m)}^{\pm}$.

\subsection{Gravitational perturbations of the Kerr black hole}

Teukolsky equations \cite{teuko} are a set of separable partial differential equations for linear fields on type-D backgrounds.  
Two of them 
describe the behavior of $\delta \Psi_0$ and $\delta \Psi_4$ for the type-D background (e.g., a perturbed Kerr black hole),
 assuming a background null tetrad with 
$k^a$ and $l^a$ aligned along the repeated PNDs. Although the connection of theses quantities with the corresponding metric perturbation 
is rather intricate \cite{Wald:1978vm}, Teukolsky equations -unlike the Regge-Wheeler and Zerilli equations for a Schwarzschild 
background- are particularly well suited to our purposes since they give the quantities needed in (\ref{split}), (\ref{ls}) and 
(\ref{ks}). \\
It was shown in \cite{wald73} that for well behaved (meaning satisfying suitable boundary conditions at the horizon and infinity)
 {\em non-stationary} black hole perturbations $\delta \Psi_0$ and 
$\delta \Psi_4$ uniquely determine each other. In particular, $\delta \Psi_0=0$ if and only if $\delta \Psi_4=0$, and this corresponds 
to a trivial perturbation. In view 
of (\ref{ks}) and (\ref{ls}), both repeated PNDs split and the perturbed metric is type I. Non-stationary perturbations are those 
relevant to the black hole stability issue and are the ones we focus on in this work. 

\subsection{Chandrasekhar algebraic special modes and  Schwarzschild's naked singularity instability} \label{asp}

In his 1984 paper \cite{asm}, Chandrasekhar dealt with the problem of finding perturbations of black holes in the Kerr-Newman 
family such that  one of $\delta \Psi_0, \delta \Psi_4$ vanishes while the other does not. 
It follows from the comments in the previous subsection that these perturbations do not satisfy appropriate 
boundary conditions at the horizon or at infinity of a Kerr-Newman black hole. 
The requirement that $\delta \Psi_0 =0$ or $\delta \Psi_4=0$ leads  to an algebraic condition 
(the vanishing of the Starobinsky constant) that gives a relation among the black hole parameters, the harmonic number of 
the perturbation and its 
frequency $\omega$ (perturbations behave as $\sim e^{i\omega t}$ for pure modes). Although Chandrasekhar's algebraic special (AS) 
modes  in the Kerr-Newman family diverge either at infinity or at the horizon of a black hole background, 
for naked singularities in the Kerr-Newman family 
 some AS are indeed very relevant, as  they satisfy appropriate boundary conditions both at infinity and at the singularity, 
and they grow exponentially with time (i.e., have a purely imaginary  $\omega$). The existence of these modes was indeed 
crucial to prove 
 the linear instability of the negative mass Schwarzschild solution 
and of the the super-extreme Reissner-Nordstr\"om 
solution (for the super-extreme Kerr solution, however, 
none of the AS modes satisfy appropriate boundary conditions and other methods 
were required to establish its linear instability). \\

The instability of the Schwarzschild solution (\ref{schwa}) with $M<0$  is due to the existence 
of a family of even/scalar  solutions of the Zerilli equation of  the form 
\begin{equation} \label{asm}
\phi_{(\ell,m)}^+ = \frac{r (r-2M)^k}{(\ell+2)(\ell-1)r + 6M} \; \exp \left( \frac{k(r-t)}{2M} \right), \;\;\; k= \frac{(\ell+2)!}{6 (\ell-2)!}.
\end{equation}
These were  found in \cite{Gleiser:2006yz}, then  recognized in \cite{Cardoso:2006bv} to agree the the AS modes in \cite{asm}.
The facts that they behave properly 
for $r \in (0,\infty)$ (keep in mind that $M<0$ and that $\ell \geq 2$ for non-stationary perturbations,
 for further details see \cite{Gleiser:2006yz}) and grow 
exponentially with time, signals the instability of the naked singularity. For the perturbations (\ref{asm}) we find
\begin{equation}
\delta \Psi_0 = 0, \;\;\; \delta \Psi_4 = \frac{6k}{M^2}\; 
(r-2M)^{k-1} \left[(\ell+2)(\ell-1) + \frac{6M}{r} \right]  \; \exp \left( \frac{k(r-t)}{2M} \right) \; Y_{-2}^{(\ell,m)}
\end{equation}
According to (\ref{ks}) and (\ref{ls}) the double PND $k^a$ will remain double whereas $l^a$ will split, the perturbed
spacetime being being of type II according to Table I. Note, however, that this can only be accomplished 
by fine tuning the perturbation to restrict to the modes (\ref{asm}), a generic perturbation will also contain 
the stable, oscillating modes, and  the PNDs will split into four, that is, to type I.

\section{Discussion}

We have found explicit formulas for the splitting of the repeated PNDs of type D spacetimes under  gravitational perturbations, 
and also   for the splitting of the repeated   eigenvalue in (\ref{ev-i}). These are given in 
equations 
(\ref{ks}) and (\ref{ls}) and  (\ref{split}).
These are observable (gauge invariant) effects of the perturbation on the background geometry that do not 
require higher than two derivatives of the metric, in contrast to differential invariants. \\
In view of (\ref{ks}), (\ref{ls}) 
and the results in \cite{wald73}, perturbed black holes within the Kerr-Newman family suffer a PND splitting to type I,  
except for stationary perturbations, which by the black hole uniqueness theorems 
are restricted to changes in the mass and/or angular momentum 
 parameters, and therefore trivially keep the type D structure. 
 This gives sense to the notion that the ``tangent vector'' $h_{ab}$ 
of a curve $g_{ab}(\epsilon)$ at a black hole metric $g_{ab}(0)$ will ``stick out''  of the set of algebraically special 
metrics. Note, however, that boundary conditions play a crucial role in these assertions: the example in Section \ref{asp} 
shows the flow of the type D Schwarzschild naked singularity to a type II  spacetime, triggered by the 
instability. This flow, however, can only be occur for fine tuned  initial conditions allowing only the (infinitely many) 
modes (\ref{asm}). A generic perturbation will contain modes other than these and will therefore 
split the two repeated PNDs into four single PNDs. \\
As a final comment, the non-analytical behavior of the PNDs in the perturbation parameter (the dominant order in the
 perturbed PNDs is $\epsilon^{1/2}$) is to be expected from the polynomial character of the PND equation and the 
confluence of the solutions as $\epsilon \to 0$. This fact was clarified in \cite{cherubini}, whose results are in  total agreement 
with ours.

\acknowledgements This work was partially funded from Grants No. PIP
11220080102479 (Conicet-Argentina), and No. Secyt-UNC 05/B498 (Universidad Nacional de C\'ordoba). Bernardo Araneda is a fellow
 of Conicet.  The {\sc grtensor} package (grtensor.org) was used to calculate (\ref{dp0s}) and (\ref{dp4s}).

\end{document}